# Signal detection algorithms for interferometric sensors with harmonic phase modulation: miscalibration of modulation parameters

**LEONID LIOKUMOVICH,[1] KONSTANTIN MURAVYOV,[1] PHILIPP SKLIAROV[1], NIKOLAI USHAKOV[1,*]**

[1]Higher school of applied physics and space technologies, Peter the Great St. Petersburg Polytechnic University, 29, Polytechnicheskaya ul., St. Petersburg, 195251, Russia.
*Corresponding author: n.ushakoff@spbstu.ru

**In the current paper, we have developed an analytical apparatus, allowing to calculate the phase error, produced by miscalibration of modulation parameters. The case of harmonic modulation is considered, the analysis is performed for cases of two parameters miscalibration: amplitude and start phase. Two demodulation algorithms are considered: conventional 4-point algorithm, based on ordinary least squares approach and previously developed 4+1 algorithm with high immunity to phase error, induced by change of target phase on demodulation interval. Predictions, given by developed analytical equations are verified by means of numeric modeling. © 2018 Optical Society of America**

http://dx.doi.org/10.1364/AO.99.099999

## 1. INTRODUCTION

Interferometric measurements necessarily require a solution to the problem of interference signal modulation, by which means the desired phase difference of the interfering optical waves carrying information about the measured signal should be demodulated from the recorded interference signal $u(t)$. A common and effective approach used in modern practical measuring devices involves the use of auxiliary modulation of the interference signal argument [1–4]. Then, in the digital domain with a sampling frequency $f_d$, demodulation is a processing of the flow of samples of the interference signal having the form

$$u_i = u(t_i) = U_0 + U_m\cos(\varphi + \psi^{(i)}),\ \psi^{(i)} = \psi(t_i),\ t_i = i/f_d, \quad (1)$$

where $\varphi$ is the target phase difference, carrying information about the measured physical quantity; $U_0$ and $U_m$ are the constant component and the amplitude of the interference signal; $\psi(t)$ is the periodic signal of the auxiliary modulation of a known form. Taking into account the technical feasibility, the frequency of the modulating signal $f_M$ and the sampling frequency are usually selected in multiples, wherein the ratio $N = f_d/f_M$ is an integer equal to the number of samples on the period of modulation.

Each sample of the measured phase is calculated from several samples of the signal given by eq. (1), further referred to as demodulation interval. The general form of expression, used to calculate the resultant phase is given by

$$\varphi_r = \operatorname{atan2}\left(\frac{a_0 u^{(0)} + a_1 u^{(1)} + \ldots + a_{Q-1} u^{(Q-1)}}{b_0 u^{(0)} + b_1 u^{(1)} + \ldots + b_{Q-1} u^{(Q-1)}}\right), \quad (2)$$

where $\varphi_r$ – calculated value of the target phase, $Q$ is the number of samples in the interval of detection, $a_q$ and $b_q$ are fixed coefficients, $u^{(q)}$ is the $q$-th sample of interval of demodulation, atan2 is the extended inverse tangent, which allows to obtain a value in the range (–π; π].

It should be noted that most methods used for phase demodulation from signals of form given by eq. (1) utilize an equation similar to eq. (2) for phase calculation (such algorithm are often referred to as linear). Earlier approaches of phase demodulation in fiber-optic interference sensors, such as widely used until nowadays phase generated carrier (PGC) [5], utilized harmonic auxiliary modulation and analogue transformations like multiplications with reference signals ($\sin(2\pi f_M t)$ and $\sin(4\pi f_M t)$) and integration. As a result of these transformations, quadrature components of a form $S=\sin(\varphi)$ and $C=\cos(\varphi)$ can be obtained, as a consequence of analytical properties of signal in eq. (1) with harmonic modulation, which can be written as $\cos[\varphi +\Psi_m \cdot \sin(2\pi f_M t)]$. However, in case of digital representation of signals and their processing, a correspondence to eqs. (1) and (2) is obtained, where components *S* and *C* are given by nominator and denominator of fraction in atan2 argument in eq. (2).

Demodulation algorithms, directly based on digital representation of signal in eq. (1), are generally considered as solutions of system of *Q* equations similar to eq. (1) under assumption of known $\Psi^{(q)}$ and $U_0$, $U_m$ and $\varphi$ being treated as unknown variables. The resulting expression for target phase $\varphi$ is reduced to the form of eq. (2), be exception of several so-called non-linear algorithms. A redundancy arises in case of *Q* > 3, leading to possibly several different solutions in case of given modulation parameters and values of *N* and *Q*. Least-squares approach or more specific methods like the one considered in [6] can be used for this purpose. Different solutions differ by $a_q$ and $b_q$ coefficient values, but lead to the same target phase $\varphi$ values in case of all parameters being static on demodulation interval and modulation parameters are exactly known. In real systems various distortions from ideal conditions lead to difference between real value $\varphi$ and calculated value $\varphi_r$. Different algorithms can have different sensitivities to different distortion mechanisms, which makes it important to investigate the properties of different demodulation algorithms.

In [6], we proposed two computational algorithms: the algorithm OLS-4 and the compensation algorithm 4+1, using harmonic phase modulation and *N*=4. For the algorithms OLS-4 and 4+1, *Q*=*N*=4 and *Q*=*N*+1=5, respectively. The explicit form of eq. (1) for these algorithms will be given below (in section 4). For these algorithms, expressions for the phase error caused by the target phase change on the detection interval were obtained and analyzed. This error $\Delta\varphi = \varphi_r - \varphi_0$ (where $\varphi_0$ – true value of phase, which is set to $\varphi$ in the center of the interval of detection) is inevitably present and increases with the amplitude of the target phase. Therefore, such a mechanism of signal distortion was chosen as a priority, and the algorithm 4+1 proposed in [6] was developed in a manner to significantly reduce $\Delta\varphi$.

However, in practice, there are other factors that lead to deviation of the calculated phase from the desired value. One of them is inaccurate calibration of the parameters of the modulating harmonic signal, because the real values of the parameters (for example, the amplitude or the initial phase) may differ from the assumed values, which are taken into account in the coefficients $a_q$ and $b_q$, which is why the phase $\varphi_r$ will be calculated with an error. The mismatch between the actual and expected signal parameters $\psi(t)$ is generally a technical problem, and it is possible to improve the calibration quality and reduce this mismatch, although the problem is compounded by the fact that in practice the mismatch can be caused by the drift of the characteristics of the modulating device during measurements. In any case, to understand the level of distortions and requirements for calibration of modulating influence parameters, as well as to compare the algorithms for resistance to such misalignment, it is necessary to have expressions to estimate the error of this type.

This issue has already been raised in a number of works dedicated to interference signal demodulation. In [7–16] the influence of this step deviation on the resulted phase $\varphi_r$ was analyzed, and also special equations, allowing to reduce the resultant error were developed for the case of linear auxiliary modulation with step $2\pi/N$. As for harmonic auxiliary modulation, the error caused by amplitude and initial phase miscalibration for a specific demodulation algorithm with fixed parameters, minimizing the impact of additive noises is investigated in [17]. Also the works [18,19] analyze the effect the modulation amplitude error and the possibility of its adjustment for the algorithm with harmonic modulation with amplitude $\pi$ and *N*=12.

The aim of this work is to evaluate the effect of amplitude and initial phase of the modulating signal miscalibration on the operation of two computational algorithms proposed by us in [6], obtaining the calculated expressions for the error of phase determination. In addition, this paper shows that the approach to the definition of $\Delta\varphi$ based on the decomposition of the structure in eq. (2) in the Taylor series with respect to small misalignment parameter, used in [6] is universal and can be used to analyze other error mechanisms, in particular errors due to inconsistency of the actual and assumed parameters of the auxiliary modulating signal.

## 2. PROBLEM STATEMENT

In general, harmonic auxiliary modulation signal in discrete representation can be written as

$$\psi_i = \psi(t_i) = \psi_m \sin(2\pi f_M t_i + \theta) = \psi_m \sin\left[\frac{2\pi}{N} \cdot i + \theta\right], \quad (3)$$

where $\psi_m$ and $\theta$ – are the amplitude and starting phase of the modulating signal. Algorithms, proposed in [6], assumed $\theta=0$. Modulation amplitude can be arbitrary, but exactly known and not be a multiple of $\pi$. In case of violation of the latter condition, the systems of equations, needed to be solved in order to find $\varphi_r$, becomes degenerate, hence, the target phase can't be found.

In case of calibration errors for amplitude and start phase of modulating signal $\psi(t)$, then the real modulation signal can be written as

$$\psi_i = \psi_m \cdot (1 + \Delta_\psi) \cdot \sin\left[\frac{2\pi}{N} \cdot i + \theta + \Delta_\theta\right], \quad (4)$$

where $\Delta_\psi$ and $\Delta_\theta$ are deviations of practical amplitude and start phase values from the assumed ones. Since in practical situations drift of $\psi(t)$ signal parameters is quasi-static, on a relatively short demodulation interval with length *Q* samples, $\Delta_\psi$ and $\Delta_\theta$ can be assumed to be constant. Moreover, during the analysis of distortions, caused by presence of $\Delta_\psi$ and $\Delta_\theta$, we'll assume the target phase $\varphi$ constant on demodulation interval as well and equal to $\varphi_0$.

In case of $\Delta_\psi = \Delta_\theta = 0$ (distortions are absent), substituting samples $u^{(q)}$ into equations, used to calculate the $\varphi_r$, and obtaining analogues of eq. (2), the structure of these equations must correspond to the form $\varphi_r = \text{atan2}(S/C)$, where *S* and *C* are equal to $\sin\varphi_0$ and $\cos\varphi_0$, providing strict equality $\varphi_r = \varphi_0$. When taking into account nonzero values of $\Delta_\psi$ and $\Delta_\theta$, the structure of equation, providing the $\varphi_r$ distorts and $\varphi_r = \text{atan2}(S_r/C_r)$, where $S_r$ and $C_r$ are different from *S* and *C*, and the phase $\varphi_r$ differs from the $\varphi_0$. Phase error is introduced as $\Delta\varphi = \varphi_r - \varphi_0$, and the goal of this work is to find the relation between $\Delta\varphi$ and $\Delta_\psi$ and $\Delta_\theta$ for phase demodulation algorithms, proposed in [6]. This will allow to analyze the influence of these mechanisms and to estimate the requirements to the systems, providing the modulating signal.

## 3. ESTIMATING PHASE ERRORS IN CASE OF DETERMINED DISTORTIONS

In [6] we considered the phase error, caused by target phase change on the demodulation interval (under assumption of linear phase change). Target signal increment $\delta$, taking place during the sampling period $1/f_d$ was chosen as a small parameter, with respect to which the Tailor expansion of $\varphi_r = \text{atan2}(S_r/C_r)$ was done. Then the phase error could be approximated by series terms, proportional to either $\delta$ or $\delta^2$ (when $\sim\delta$ term was equal to zero). This approach turns out to be universal and applicable for distortion of other types, including the ones, considered in this paper. Let the considered distortion be characterized by parameter $\Delta \ll 1$, and the condition $\Delta\varphi \ll 1$ is valid as well. Then $\Delta\varphi$ can be approximated by

$$\Delta\varphi \approx \frac{d}{d\Delta}\mathrm{atan2}\left(\frac{S_\mathrm{r}}{C_\mathrm{r}}\right)\bigg|_{\Delta=0}\cdot\Delta = \Delta\frac{S'C - C'S}{C^2 + S^2}, \quad (5)$$

where $S'$ and $C'$ are first derivatives of $S_\mathrm{r}$ and $C_\mathrm{r}$ with respect to $\Delta$ in case of $\Delta=0$. If $S'C=C'S$, then eq. (5) turns to zero, in this case one must consider $\Delta\varphi$ as a next series term, i.e.

$$\Delta\varphi \approx \frac{1}{2}\frac{d^2}{d\Delta^2}\mathrm{atan2}\left(\frac{S_\mathrm{r}}{C_\mathrm{r}}\right)\bigg|_{\Delta=0}\cdot\Delta^2 =$$
$$\frac{1}{2}\Delta^2\frac{(S''C - C''S)(C^2 + S^2) - 2(CC' + SS')(S'C - C'S)}{(C^2 + S^2)^2} =$$
$$= \frac{1}{2}\Delta^2\frac{S''C - C''S}{C^2 + S^2}, \quad (6)$$

where $S''$ and $C''$ are second derivatives of $S_\mathrm{r}$ and $C_\mathrm{r}$ with respect to $\Delta$ in case of $\Delta=0$. If this quantity is equal to zero as well, further terms of Tailor series must be considered.

For algorithms with linear structure like eq. (2), $S'$, $C'$, $S''$ and $C''$ will have a form

$$S' = \sum_{q=0}^{Q-1}\left[a_q\cdot\frac{du_\mathrm{r}^{(q)}}{d\Delta}\bigg|_{\Delta=0}\right], \quad C' = \sum_{q=0}^{Q-1}\left[b_q\cdot\frac{du_\mathrm{r}^{(q)}}{d\Delta}\bigg|_{\Delta=0}\right],$$
$$S'' = \sum_{q=0}^{Q-1}\left[a_q\cdot\frac{d^2u_\mathrm{r}^{(q)}}{d\Delta^2}\bigg|_{\Delta=0}\right], \quad C'' = \sum_{q=0}^{Q-1}\left[b_q\cdot\frac{d^2u_\mathrm{r}^{(q)}}{d\Delta^2}\bigg|_{\Delta=0}\right]. \quad (7)$$

Therefore, in order to find the error estimates for a given algorithm, derivatives of analytical representations of signal samples $u^{(q)}$ must be calculated, then substituted into eq. (7) together with coefficients $a_q$ and $b_q$ and then plugged into eqs. (5), (6) or similar equations for higher-order terms.

## 4. PHASE ERRORS

Let us consider algorithms, proposed in [6], implying modulating signal to have a form of eq. (3) and $N=4$, $\theta=0$. In this case samples of interference signal can be written as follows

$$u_\mathrm{r}^{(0)} = u_\mathrm{r}^{(4)} = U_0 + U_\mathrm{m}\cos\left[\varphi_0 + \psi_\mathrm{m}(1+\Delta_\psi)\cdot\sin(\Delta_\theta)\right],$$
$$u_\mathrm{r}^{(1)} = U_0 + U_\mathrm{m}\cos\left[\varphi_0 + \psi_\mathrm{m}(1+\Delta_\psi)\cdot\cos(\Delta_\theta)\right],$$
$$u_\mathrm{r}^{(2)} = U_0 + U_\mathrm{m}\cos\left[\varphi_0 - \psi_\mathrm{m}(1+\Delta_\psi)\cdot\sin(\Delta_\theta)\right], \quad (8)$$
$$u_\mathrm{r}^{(3)} = U_0 + U_\mathrm{m}\cos\left[\varphi_0 - \psi_\mathrm{m}(1+\Delta_\psi)\cdot\cos(\Delta_\theta)\right].$$

In OLS-4 algorithm, which was developed according to least-squares method [1] for eq. (3) in case of $Q=N=4$, $\theta=0$, $u^{(0)}$, $u^{(1)}$, $u^{(2)}$ and $u^{(3)}$ signal samples from a single period are used for one phase sample calculation, which is expressed as

$$\varphi_\mathrm{r} = -\mathrm{atan2}\left[\frac{1-\cos(\psi_\mathrm{m})}{\sin(\psi_\mathrm{m})}\cdot\frac{u^{(1)} - u^{(3)}}{u^{(0)} - u^{(1)} + u^{(2)} - u^{(3)}}\right]. \quad (9)$$

It should be noted that $[1-\cos\psi_\mathrm{m}]/\sin\psi_\mathrm{m}$ factor must be treated as a part of the nominator in order to ensure correct calculation of atan2 function.

In 4+1 algorithm, five signal samples are used for phase sample calculation, which means that $u^{(4)}$ (in our case equal to $u^{(0)}$) is included for computation. Thus $Q=N+1=5$, leading to $(N+1)$ denotation of all algorithms, using such sample set [3,7,8]. This algorithm was developed according to the approach, proposed in [8], ensuring the $\Delta\varphi$ components, proportional to $\delta$ and $\delta^2$ are equal zero. The phase sample is calculated by formula [6]

$$\varphi_\mathrm{r} = -\mathrm{atan2}\left[\frac{1-\cos(\psi_\mathrm{m})}{\sin(\psi_\mathrm{m})}\cdot\frac{2u^{(0)} + 4u^{(1)} - 4u^{(3)} - 2u^{(4)}}{u^{(0)} - 4u^{(1)} + 6u^{(2)} - 4u^{(3)} + u^{(4)}}\right] \quad (10)$$

Again, $[1-\cos\psi_\mathrm{m}]/\sin\psi_\mathrm{m}$ factor must be treated as a part of the nominator in order to ensure correct calculation of atan2 function.

### A. Miscalibration of modulation amplitude

Let us consider the phase error, induced by miscalibration of actual and assumed modulation amplitude. Derivatives of equations in eq. (8) with respect to $\Delta_\psi$ in case of $\Delta_\theta=0$ will be

$$\frac{du_\mathrm{r}^{(0)}}{d\Delta_\psi}\bigg|_{\Delta_\psi=0} = 0, \quad \frac{du_\mathrm{r}^{(2)}}{d\Delta_\psi}\bigg|_{\Delta_\psi=0} = 0, \quad \frac{du_\mathrm{r}^{(4)}}{d\Delta_\psi}\bigg|_{\Delta_\psi=0} = 0$$
$$\frac{du_\mathrm{r}^{(1)}}{d\Delta_\psi}\bigg|_{\Delta_\psi=0} = -U_\mathrm{m}\sin\left[\varphi_0 + \psi_\mathrm{m}\right]\cdot\psi_\mathrm{m}, \quad (11)$$
$$\frac{du_\mathrm{r}^{(3)}}{d\Delta_\psi}\bigg|_{\Delta_\psi=0} = U_\mathrm{m}\sin\left[\varphi_0 - \psi_\mathrm{m}\right]\cdot\psi_\mathrm{m}.$$

Substituting eq. (11) to eq. (7) and then to eq. (5), an estimate of phase error can be obtained

$$\Delta\varphi \approx -\frac{\Delta_\psi}{2}\cdot\psi_\mathrm{m}\cdot\frac{\sin(2\varphi_0)}{\sin(\psi_\mathrm{m})}. \quad (12)$$

Turns out, eq. (12) is valid for both OLS-4 and 4+1 algorithms.

It follows from eq. (12) that $\Delta\varphi$ is periodical with respect to $\varphi_0$ and is proportional to $\mathrm{sinc}(\psi_\mathrm{m})$. Also, there is no growth of error in case of $\psi_\mathrm{m}=0$, even though this value is inadmissible due to the algorithm performance. Rapid growth of higher-orders of magnitude error components could be expected in this case, however, calculations show that this is not the case and eq. (12) describes the behavior of the error well.

For interference sensors $\varphi_0$ can be arbitrary, since it varies according to both target perturbation and parasitic quasi-static drift of parameters. Since above it is assumed that $\varphi_0$ is constant on a demodulation interval, the possibility of its change must be treated as from one demodulation interval to another. Hence, it follows from eq. (12) that the worst case of start phase is $\sin(2\varphi_0)=\pm 1$ and least error in this case is achieved when $\psi_\mathrm{m}=\pi/2+k\pi$ and is $\Delta\varphi_\mathrm{min} = \pm\Delta_\psi/2$. As could be expected, the error will be significant in case of $\psi_\mathrm{m}$ close to multiple of $\pi$. General form of dependency given by eq. (12) in case of $\sin 2\varphi_0 = 1$ is illustrated by curve 1 in figure 1, the other bound of maximal absolute value of the error for $\sin 2\varphi_0 = -1$ is illustrated by the inversed curve.

Due to evident technical reasons, the smallest auxiliary modulation amplitudes, providing given error value are of the major interest. For example, setting the acceptable error limit to $2|\Delta\varphi_\mathrm{min}|$, then, according to eq. (12), amplitudes $\psi_\mathrm{m}$ must fall into the interval (0, 1.9] radians. However, in practice, the lower limit must be greater, according to other types of distortions and noises, which can affect the algorithm performance in proximity of this range bounds.

It should be noted that for $\psi_m$ in vicinity of $k\pi$ ($k$ – integer), rapid growth of the $\Delta\varphi$ error takes place, and, moreover, the eq. (12) itself becomes inadequate. This is so because for these values of $\psi_m$ higher order terms in Tailor expansion of $\varphi_r$ with respect to $\Delta_\psi$ become comparable to the first term and approximation by eq. (12) becomes incorrect. Considering this expansion in more detail, one can obtain analytical conditions for $\psi_m$ value, allowing to use eq. (12) in case of a given requirements to the error estimation accuracy, however, this falls out of the scope of this article.

### B. Miscalibration of modulation start phase

Now let us consider the error, induced by the modulation start phase being different from zero in case of $\Delta_\psi = 0$. Derivatives of samples in eq. (8) with respect to $\Delta_\theta$ are

$$\left.\frac{du_r^{(0)}}{d\Delta_\theta}\right|_{\Delta_\theta=0} = -U_m \cdot \sin(\varphi_0) \cdot \psi_m, \quad \left.\frac{du_r^{(1)}}{d\Delta_\theta}\right|_{\Delta_\theta=0} = 0,$$

$$\left.\frac{du_r^{(2)}}{d\Delta_\theta}\right|_{\Delta_\theta=0} = U_m \sin(\varphi_0) \cdot \psi_m,$$

$$\left.\frac{du_r^{(3)}}{d\Delta_\theta}\right|_{\Delta_\theta=0} = 0, \quad \left.\frac{du_r^{(4)}}{d\Delta_\theta}\right|_{\Delta_\theta=0} = -U_m \cdot \sin(\varphi_0) \cdot \psi_m.$$

(13)

Substituting eq. (13) into eq. (7) and eq. (5), for OLS-4 yields zero for the first order of magnitude error component. Therefore, one needs to consider the second order of magnitude terms. Second derivatives of samples in eq. (8) with respect to $\Delta_\theta$ are

$$\left.\frac{d^2 u_r^{(0)}}{d\Delta_\theta^2}\right|_{\Delta_\theta=0} = -U_m \cdot \cos(\varphi_0) \cdot \psi_m^2,$$

$$\left.\frac{d^2 u_r^{(1)}}{d\Delta_\theta^2}\right|_{\Delta_\theta=0} = U_m \cdot \sin(\varphi_0 + \psi_m) \cdot \psi_m,$$

$$\left.\frac{d^2 u_r^{(2)}}{d\Delta_\theta^2}\right|_{\Delta_\theta=0} = -U_m \cdot \cos(\varphi_0) \cdot \psi_m^2,$$

$$\left.\frac{d^2 u_r^{(3)}}{d\Delta_\theta^2}\right|_{\Delta_\theta=0} = -U_m \cdot \sin(\varphi_0 - \psi_m) \cdot \psi_m.$$

(14)

Substituting eq. (14) into eq. (7) and eq. (6), the following error estimate is obtained

$$\Delta\varphi \approx \frac{\Delta_\theta^2}{4} \sin(2\varphi_0) \cdot \left[\frac{\psi_m}{\sin(\psi_m)} + \frac{\psi_m^2}{1-\cos(\psi_m)}\right].$$

(15)

In such a manner, OLS-4 algorithm supresses modulation start phase miscalibration, removing the first order of magnitude error component.

It can be seen that in this case $\Delta\varphi$ again has periodical dependence with respect to $\varphi_0$, however, dependence on the modulation amplitude is much more complex than in eq. (12). It can be seen from eq. (15) that the phase error for OLS-4 algorithm grows rapidly in the vicinity of inadmissible values of $\psi_m$ multiples of $\pi$. However, in case of $\psi_m=0$ there is no growth of error even though this value is inadmissible due to the algorithm performance. Again, rapid growth of higher-orders of magnitude error components could be expected, however, it is not the case. The worst case in terms of $\varphi_0$ is $\sin(2\varphi_0)=\pm 1$, resulting in $\Delta\varphi_{min}=\pm 3\Delta_\theta^2/4$. Therefore, in analogy with the criterion above, implying the error $|\Delta\varphi| < 2|\Delta\varphi_{min}|$, modulation amplitude values can be limited to a range (0, 2.25] radians according to eq. (15). However, as already mentioned above, the lower limit must be greater, according to other types of distortions and noises. General form of dependency given by eq. (15) in case of $\sin 2\varphi_0 = 1$ is illustrated by curve 2 in figure 1, the other bound of maximal absolute value of the error for $\sin 2\varphi_0 = -1$ is illustrated by the inversed curve.

Now let us consider 4+1 algorithm. Substituting eq. (13) into eq. (7) and eq. (5), the following error estimate is obtained

$$\Delta\varphi = -\frac{\Delta_\theta}{4} \cdot \psi_m \cdot \frac{1-\cos(2\varphi_0)}{1-\cos(\psi_m)}.$$

(16)

It follows from eq. (16) that 4+1 algorithm is less robust in terms of modulation start phase miscalibration compared to OLS-4 algorithm. However, as was shown in [6], 4+1 algorithm is less susceptible to error, caused by target phase variation on the demodulation interval.

Again, $\Delta\varphi$ varies periodically with respect to $\varphi_0$ and aperiodically with respect to $\psi_m$. It can be seen from eq. (16) that the worst case corresponds to $\varphi_0=\pm\pi/2$, in which case the error is $\Delta\varphi=-\Delta_\theta\psi_m/[1-\cos\psi_m]$. It also should be mentioned that the error is biased and always negative, increasing in case of modulation amplitude close to multiple of $2\pi$. Minimal error is yielded for modulation amplitude 2.331 radians and is equal to $\Delta\varphi_{min}=-0.69\Delta_\theta$. Formulating a criterion $|\Delta\varphi|<2|\Delta\varphi_{min}|$, one obtains the following limitation on the modulation amplitude: $0.765$ rad $< \psi_m < 4.18$ rad. General form of dependency given by eq. (16) in case of $\sin 2\varphi_0 = 1$ is illustrated by curve 1 in figure 1, the other bound of maximal absolute value of the error for $\sin 2\varphi_0 = -1$ is illustrated by the inversed curve.

However, even though the eq. (16) itself doesn't predict error growth in case of $\psi_m$ close to $\pi$, numeric modeling, described below, demonstrates significant phase errors in case of $\psi_m \sim \pi+2\pi k$, which is related to growth of higher order error components in proximity of these points. Therefore, modulation amplitude must not exceed $\pi$.

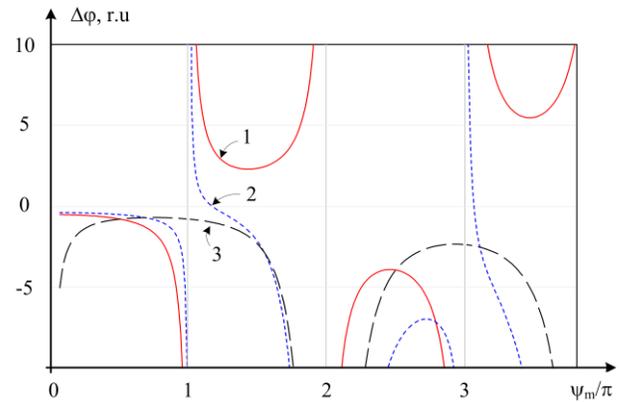

Figure 1. Demonstration of eqs. (12), (15) and (16) behavior with respect to modulation amplitude normalized by $\pi$ for worst cases of $\varphi_0$. Curve 1 – relation $\Delta\varphi/\Delta_\psi$ in case of $\varphi_0=\pi/4$, calculated according to eq. (12); curve 2 – relation $\Delta\varphi/(2\Delta_\theta^2)$ in case of $\varphi_0=\pi/4$, calculated according to eq. (15); curve 3 – relation $\Delta\varphi/\Delta_\theta$ in case of $\varphi_0=\pi/4$, calculated according to eq. (16).

## 5. INFLUENCE OF SIMULTANEOUS MISCALIBRATION OF MODULATION AMPLITUDE AND STARTING PHASE

The analysis of errors caused by the considered distortions should take into account simultaneous presence of different types of distortions that can take place in real systems, such as in particular, simultaneous presence of nonzero $\Delta_\psi$ and $\Delta_\theta$. Then, within the given approach to the error analysis, it is necessary to consider the expansion of $\Delta\varphi$ in Taylor series, as functions of two variables. The first order of magnitude of the series correspond to individual dependencies on each parameter, considered above. Along with these terms and terms proportional to greater powers of $\Delta_\psi$ and $\Delta_\theta$, the term, proportional to the cross product of different distortion parameters, i.e. $\Delta_\psi \cdot \Delta_\theta$ will be present as well.

However, for the considered algorithms, the cross-component can be omitted. Indeed, if $\Delta_\psi$ and $\Delta_\theta$ are quantities of one order of magnitude, for OLS-4 algorithm the phase error will be determined mainly by the eq. (12), since $\Delta_\psi \cdot \Delta_\theta \ll \Delta_\psi$, and the cross-component can be neglected, as well as other components of the second order of magnitude. Moreover, eq. (12) will dominate in case of $\Delta_\psi \gg \Delta_\theta$. In case of $\Delta_\psi \gg \Delta_\theta$, components $\Delta_\psi$ and $\Delta_\theta^2$ will be comparable and the error will be determined by sum of eqs. (12) and (15). In turn, $\Delta_\psi \cdot \Delta_\theta$ and $\Delta_\psi^2$ are negligible compared to $\Delta_\psi$ and $\Delta_\theta^2$, enabling one not to take the cross-component into account.

For the 4+1 algorithm, according to eqs. (12) and (16), the errors are determined by the terms with first power of $\Delta_\psi$ and $\Delta_\theta$, so the cross-component $\Delta_\psi \cdot \Delta_\theta$ can always be neglected. In general, the error estimate can be considered as the sum of the eqs. (12) and (16). Although if $\Delta_\psi \gg \Delta_\theta$, the term (12) (and hence, the value of $\Delta_\psi$) will determine the resultant error. Otherwise, when $\Delta_\theta \gg \Delta_\psi$ – the main contribution to the error will be given by eq. (16) and depend mainly on $\Delta_\theta$.

As a result, it can be concluded that in case of simultaneous presence of disorder in amplitude and phase, the error in general should be considered as the sum of eqs. (12) and (15) for the algorithm OLS-4, or eqs. (12) and (16) for the 4+1 algorithm, although at certain conditions one of the components may dominate. Cross component with a multiplier with $\Delta_\psi \cdot \Delta_\theta$ can be ignored, however, for $\psi_m$ in proximity to inadmissible values, the cross-component may not be negligible and have a significant impact on the phase error.

The analysis above has supposed that the target phase $\varphi$ is constant on the demodulation interval. If it changes, it leads to a phase detection error [20], which was considered in detail in [6]. Moreover, this type of the error was considered as a priority, in relation to the rest and was the main motivation for creating the 4+1 algorithm. At the same time, in real systems, the factor of target phase changing can and will take place simultaneously by miscalibrating the parameters of the modulating signal. Therefore, it makes sense to consider how these error mechanisms interact with each other and affect the compensation properties of the algorithms.

In [6] it was shown that for the OLS-4 the phase error $\Delta\varphi$ caused by the change $\varphi$ on the demodulation interval, is proportional to $\delta$ – the first derivative of the target signal with respect to sample number. If the modulation amplitude is miscalibrated, then the error will contain the component described by the eq. (12), as well as a cross-component proportional to the product $\delta \cdot \Delta_\psi$. Since it is assumed that the miscalibration is slight, regardless of the ratio of values $\Delta_\psi$ and $\delta$, $\delta \gg \delta \cdot \Delta_\psi$ and $\Delta_\psi \gg \delta \cdot \Delta_\psi$ will hold. Thus, the cross-component will be negligible, and these error sources can be taken into account independently, summing the error estimates for each of them. In case of presence of phase change on demodulation interval and modulation start phase miscalibration, the situation is similar. In case of any ratio of $\delta$ and $\Delta_\theta$, $\delta \gg \delta \cdot \Delta_\theta$ will be true. Therefore, the error is determined by the sum of the expressions for the phase change caused error and eq. (15), although depending on the ratio of values $\delta$ and $\Delta_\theta^2$ one of the components may prevail. The cross-component again can be neglected.

Now let us consider the 4+1 algorithm. When developing this algorithm was laid down the condition that the components of the error, proportional $\delta$ and $\delta^2$, was zero. Accordingly, the error $\Delta\varphi$ caused by changing of the target phase on the demodulation interval, for a given algorithm as a first approximation, be proportional to $\gamma$ – the second derivative of the target signal with respect to the sample number of while $\gamma \ll \delta$. Accordingly, in case of the modulating signal parameters miscalibration, the sum of eqs. (12) and (16) and the cross-components, proportional to the $\delta \cdot \Delta_\psi$ and $\delta \cdot \Delta_\theta$ will be added to the initial error. However, since $\Delta_\psi \gg \delta \cdot \Delta_\psi$ and $\Delta_\theta \gg \delta \cdot \Delta_\theta$, the components of the error (12) or (16) will dominate the cross-components. With regard to the properties of the algorithm to suppress the error caused by the phase change, they either will not change (if $\delta^2 \gg \delta \cdot \Delta_\psi$ and $\delta^2 \gg \delta \cdot \Delta_\theta$ and the error studied in [6] does not change), or will no longer be playing the role (if $\Delta_\psi \gg \gamma$ and $\Delta_\theta \gg \gamma$, the main error will be related with $\Delta_\psi$ and $\Delta_\psi$).

## 6. SIMULATION RESULTS

The above analysis is based on analytical expressions, which were derived using certain conditions and approximations, therefore, additional verification of the results is desirable. In this regard, two circumstances should be noted. First, in real optical fiber interferometric schemes, provided that the polarizations of interfering beams match (by use of PM elements or schemes with Faraday mirrors), the acquired signal corresponds to the form of eq. (1) with high accuracy. Secondly, the distortions of the measured phase analyzed in this article do not arise in the interference scheme or in the blocks of signal formation and registration, but in the process of phase demodulation by the eqs. (9) or (10) due to the fact that these expressions were derived under the assumption $\Delta_\psi = \Delta_\theta = 0$. Given these circumstances, it is appropriate to use numeric simulation for verification of the obtained analytical results, including the direct calculation of the interference signal samples and their demodulation according to the eqs. (9) and (10).

Since this analysis does not include noise, in order to make a direct numerical calculation of the error $\Delta\varphi$ for fixed parameters of the signal $\varphi_0$, $\psi_m$, $\Delta_\psi$ and $\Delta_\theta$, it is sufficient to calculate 4 (for the algorithm "OLS-4") or 5 (for the 4+1 algorithm) samples of the form (8), then calculate $\varphi_r$ according to eqs. (9) or (10), and then find the difference $\Delta\varphi = \varphi_r - \varphi_0$. Such relatively simple calculations not only allow us to check the correctness of the eqs. (12), (15) and (16), but also to consider the error in case these expressions are not consistent, when the conditions of the small parameter of the miscalibration or the smallness of the error itself are not met. Numerical simulations also show additional $\psi_m$ values, providing singularities that were not obvious when obtaining analytical expressions for error estimates and do not follow from the analysis of these expressions.

However, the complexity of presenting the results is due to a large number of parameters. Along with the parameters of miscalibration, which for brevity can be considered only individually, there are two important parameters exerting great influence on the error – $\varphi_0$ and $\psi_m$. With a focus on priority issues for building practical systems, we present error dependencies on the modulation amplitude $\psi_m$, treating $\Delta_\psi$, $\Delta_\theta$ and $\varphi_0$ as parameters of these dependencies. For the same practical reasons, the range of amplitudes from small values to the first points of singular growth of error is the most relevant for the analysis. Greater amplitudes are less convenient to use and increase the error associated with $\Delta_\theta$, as can be seen from figure 1.

Consider the case of the presence of only the miscalibration of the amplitude of the modulating signal. First of all, we note that the simulation results for both algorithms under consideration were absolutely identical for all values of the parameters, including when the miscalibration does not meet the conditions of smallness. In figure 2

(a) an example of the calculations of the error $\Delta\varphi$ for small detuning $\Delta_\psi = 0.01$ with a target phase $\varphi_0 = \pi/4$, i.e. for the phase with the highest value $|\Delta\varphi|$ is shown. The figure shows a very good agreement of the modeling with the analytical equation (12). The results, obtained for $\varphi$ different from $\pi/4$, correspond very well with the analytical estimation as well. In figure 2, (b) an example of calculation for a sufficiently significant miscalibration $\Delta_\psi = 0.1$ and $\varphi_0 = \pi/4$ is shown. It can be seen from the figure that a noticeable discrepancy arises for the region of error growth when the amplitude becomes comparable to the singularity point $\pi$. Within the above-mentioned reasonable range of amplitudes 0-1.9 radians, the discrepancy is insignificant and eq. (12) can be confidently used to estimate the error. Calculations for $\Delta_\psi = 0.1$ showed that in the region $\psi_m > 2$ radians for some phases $\varphi$ different from $\pi/4$ one can obtain a more noticeable difference between the simulation result and the analytical estimate, but even though, the values $|\Delta\varphi|$ are less than the ones shown in figure 2 (b) (for the analytical evaluation the latter follows from eq. (12) ). At $\psi_m < 2$ radian for any phases $\varphi_0$ the discrepancy between the simulation and the analytical dependence is hardly noticeable. Thus, we can conclude that eq. (12) is a good estimate for the error $\Delta\varphi$ at sufficiently small $\Delta_\psi$, and even for $\Delta_\psi = 0.1$ this estimate is very good at $\psi_m < 2$ radian.

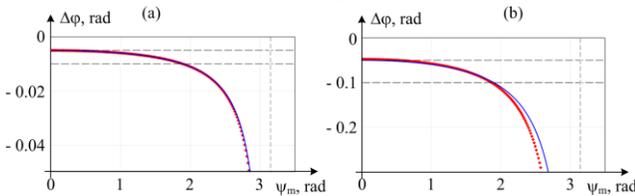

Figure 2. Modeling results of phase error $\Delta\varphi$ in case of modulation amplitude miscalibration, $\varphi_0 = \pi/4$. (a) $\Delta_\psi = 0.01$, (b) $\Delta_\psi = 0.1$. Solid line – analytical estimate according to eq. (12); points – modeling. Vertical dashed line – singularity point $\psi_m = \pi$; horizontal dashed lines show levels $|\varphi_{min}| = \Delta_\psi/2$ and $2|\varphi_{min}| = \Delta_\psi$.

Next, we give examples of calculations for the case when there is a miscalibration of the initial phase of modulation in case of the algorithm "OLS-4". In figure 3 (a), the results of calculations for $\Delta_\theta = 0.1$ and the phase $\varphi_0 = \pi/4$ corresponding to the case of the maximum error value according to eq. (15) are shown. Despite the significant level of miscalibration, the results of modeling and analytical evaluation are very close, which is due to the suppression of the error component in this algorithm proportional to the first degree of $\Delta_\theta$. When $\Delta_\theta < 0.1$ discrepancy is even smaller. A noticeable discrepancy can be seen if the miscalibration is significantly increased, for example, to $\Delta_\theta = 0.3$ (strictly, in this case, the condition $\Delta_\theta \ll 1$ doesn't hold no longer), as shown in figure 3 (b). However, even in this case, the difference between simulation result and calculation by eq. (15) are negligible in the context of assessing the value $\Delta\varphi$. The case of phase $\varphi$ different from $\pi/4$ was also studied in the simulation. It was shown that as in the case of calculation based on eq. (15), the simulation shows that at $\varphi_0 \neq \pi/4$ the error level $|\Delta\varphi|$ decreases.

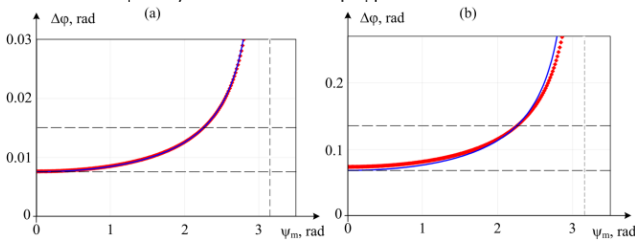

Figure 3. Modeling results of phase error $\Delta\varphi$ for OLS-4 algorithm in case of modulation start phase miscalibration, $\varphi_0 = \pi/4$. (a) $\Delta_\theta = 0.1$, (b) $\Delta_\theta = 0.3$. Solid line – analytical estimate according to eq. (15); points – modeling. Vertical dashed line – singularity point $\psi_m = \pi$; horizontal dashed lines show levels $|\varphi_{min}| = 3/4 \cdot \Delta_\theta^2$ and $2|\varphi_{min}| = 1.5\Delta_\theta^2$.

And finally, let us consider the modeling results for 4+1 algorithm for a nonzero $\theta$. In figure 4 (a) the error in case of $\Delta_\theta = 0.05$ and phase $\varphi_0 = \pi/2$, which is the worst for the analytical equation (16) is shown. A similar result is obtained for a significant miscalibration $\Delta_\theta = 0.1$, shown in figure 4 (b). Both figures show that at $\psi_m = 2.331$ radian the above-mentioned minimum $|\Delta\varphi_{min}| = 0.69\,\Delta_\theta$ is achieved, and at lower amplitudes there is a very good agreement between the simulation and analytical results given by eq. (16). However, the simulation shows the presence of a singularity point $\psi_m = \pi$, which isn't described by eq. (16). Formally, this is due to the fact that at this point (as for other amplitudes equal to $(2k+1)\pi$ ) there is a feature in the next term of the Taylor expansion of the error $\Delta\varphi$ and hence, this term can't be discarded near these points. It follows from such mismatch that the upper limit of the modulating signal amplitude should not exceed the value $\pi$, so the previously mentioned range of optimal modulation amplitudes 0.765 – 4.18 radian can be replaced by 0.765 – 3 radian.

In this case, the question of how the real error behaves in the case of $\varphi_0 \neq \pi/2$ was also studied. According to eq. (16) in this case $|\Delta\varphi|$ will necessarily be lower than in case $\varphi_0 \neq \pi/2$, but a more accurate result based on the simulation showed the possibility of violation of this rule. In particular, with a slight positive bias $\varphi_0 = d + \pi/2$ there is a change in the behavior of error in the neighborhood of $\psi_m = \pi$. At $d \approx 0.08$, the area of the singularity is actually compressed to the point $\psi_m = \pi$ and the deviation of the modeling results from eq. (16) virtually disappears, as can be seen in figure 4 (c). In case of the offset $d$ in the range of 0.1 – 0.4, the absolute values of $\Delta\varphi$ increase as the modulation amplitude approaches $\pi$ and the magnitude of the error can be greater than $|\Delta\varphi_{min}|$, although only slightly. An example of such a case is shown in figure 4 (d). In such a situation, to avoid the increase of the error and a significant discrepancy in the simulation result with eq. (16), as before, it is desirable to use $\psi_m < 3$ radian. We can conclude that when modulation amplitude of less than 3 radians, the estimate eq. (16) is in a good agreement with the simulation, especially if the condition $\Delta_\theta \ll 1$ holds.

Finally, let us present an example of calculations for the case when both parameters of modulation have a miscalibration. Figure 5 (a) shows calculated phase error for the algorithm OLS-4 in case of $\Delta_\psi = 0.2$, $\Delta_\theta = 0.01$ and $\varphi_0 = \pi/4$. Given that this algorithm suppresses the error of the first order of smallness of $\Delta_\theta$, the miscalibration parameters are chosen unequal, so that the results differ significantly from the ones calculated in case of only one of the parameters' miscalibrations present. It is easy to see good agreement between the results of modeling and error estimations based on summation of eqs. (12) and (15).

Figure 5b shows an example of the calculation for the 4+1 algorithm when $\Delta_\psi = \Delta_\theta = 0.05$ and $\varphi_0 = 1.86$. Again, in this case, the parameters are such that the calculation results differ significantly from the ones calculated in case of only one of the parameters' miscalibrations present. In this case, there are areas of noticeable differences between modeling and analytical evaluation, which, however, is explained by the above feature at $\psi_m = \pi$, which isn't described by eq. (16).

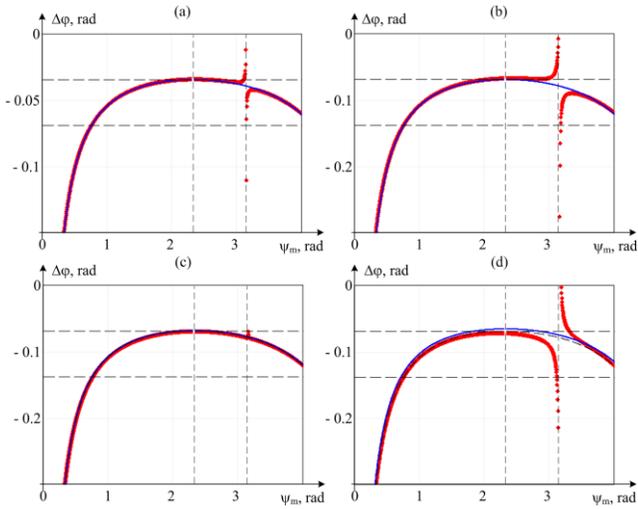

Figure 4. Modeling results of phase error $\Delta\varphi$ for 4+1 algorithm in case of modulation start phase miscalibration. (a) $\Delta_\theta = 0.05$, $\varphi_0 = \pi/2$; (b) $\Delta_\theta = 0.1$, $\varphi_0 = \pi/2$; (c) $\Delta_\theta = 0.1$, $\varphi_0 = \pi/2+0.08$; (d) $\Delta_\theta = 0.1$, $\varphi_0 = 1.8$. Solid line – analytical estimate according to eq. (16) for $\varphi_0 = \pi/2$; points – modeling. Vertical dashed lines – singularity point $\psi_m = \pi$ and case of minimal error $\varphi_0 = 2.331$; horizontal dashed lines show levels $|\varphi_{min}| = 0.69\Delta_\theta$ and $2|\varphi_{min}| = 1.38\Delta_\theta$.

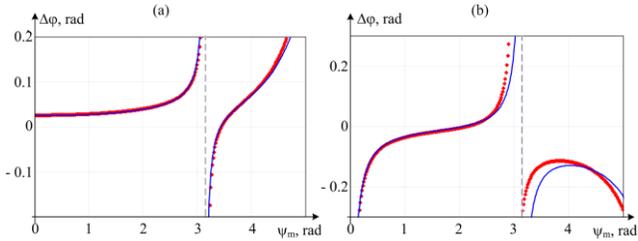

Figure 5. Modeling results of phase error $\Delta\varphi$ in case of modulation amplitude and start phase miscalibration. (a) OLS-4 algorithm, $\Delta_\psi = 0.01$, $\Delta_\theta = 0.2$, $\varphi_0 = \pi/4$; (b) 4+1 algorithm, $\Delta_\psi = \Delta_\theta = 0.05$, $\varphi_0 = 1.86$. Solid line – analytical estimate according to sum of eqs. (12) and (15) in (a) and sum of eqs. (12) and (16) in (b); points – modeling. Vertical dashed line – singularity point $\psi_m = \pi$.

The given examples show that the presented simple method of modeling interference signal samples and their processing by the given algorithm allows to analyze the actual error value $\Delta\varphi$ independent of the approximations on which the above-obtained analytical expressions are based. At the same time, based on the results of the study of the simulation results, it can be concluded that for an approximate description of the error $\Delta\varphi$, relatively simple analytical eqs. (12), (15) and (16) are adequate, given that in the latter case the amplitude range of less than 3 radians should be considered.

Given the fact that any values of $\varphi_0$ can be observed in fiber interferometers, for approximate estimates of the error level one can use estimates obtained from eqs. (12), (15) and (16) in case of the worst-case in terms of $\varphi_0$ value

$$\Delta\varphi \approx -\frac{\Delta_\psi}{2} \cdot \frac{\psi_m}{\sin(\psi_m)}.$$

$$\Delta\varphi \approx \frac{\Delta_\theta^2}{4} \cdot \left[\frac{\psi_m}{\sin(\psi_m)} + \frac{\psi_m^2}{1-\cos(\psi_m)}\right].$$

$$\Delta\varphi \approx -\frac{\Delta_\theta}{2} \cdot \frac{\psi_m}{1-\cos(\psi_m)}.$$

(17)

The first expression corresponds to the estimation of the distortion due to the amplitude miscalibration for both algorithms. The second is for miscalibration of the initial phase and the algorithm OLS-4, the third one is for miscalibration of the initial phase and 4+1 algorithm.

It should be noted that in the above reasoning the consistency of the modeling results and analytical expressions was estimated by visual coincidence or discrepancy of the calculated dependencies.

## 7. CONCLUSION

In this paper, we have developed an analytical apparatus, allowing to take into account the influence of modulation parameters miscalibration on the error of the measured phase.

Additional study, based on numeric modeling of the modulation miscalibration effects, proved the adequacy of our analytical derivations and the assumptions made. Although, it was also shown that in case of modulation start phase miscalibration there is a feature in performance of 4+1 algorithm, limiting the range of possible $\psi_m$ values.

So, according to [6] and this article, most of the phase error mechanisms are considered and, concerning the performance of 4+1 algorithm, the only remaining point is to prove that the performance of 4+1 algorithm in presence of noise isn't too bad compared to other algorithms. However, such analysis falls out of the scope of this article.

Although OLS-4 algorithm is less susceptible to modulation start phase miscalibration, 4+1 algorithm is also robust to this error source and doesn't demonstrate critical behavior in case of $\theta \neq 0$. Anyway, to our opinion, the change of the target phase on the demodulation interval is the most critical error source, since it can't be limited by any technical means, in contrast to modulation parameters miscalibration $\Delta_\psi$ and $\Delta_\theta$. Taking into account that 4+1 algorithm is much more robust to target phase change (as considered in [6]), this algorithm occurs to be preferable for practical use in demodulation systems for fiber-optic interferometric sensors, utilizing harmonic auxiliary phase modulation.

Finally, let us note that the approach, used in [6] for phase error $\Delta\varphi$ estimation in case of target phase $\varphi$ variation and in the current article for perturbation, caused by modulation parameter miscalibration, can be applied to any linear demodulation algorithm, including the ones with non-harmonic auxiliary modulation.

**Funding Information.** Ministry of Education and Science of the Russian Federation in terms of FTP "Research and development on priority trends of Russian scientific-technological complex evolvement in 2014-2020 years (agreement # 14.578.21.0211, agreement unique identifier RFMEFI57816X0211)".